\documentstyle[12pt,fleqn]{article}
\def\singlespace {\smallskipamount=3.75pt plus1pt minus1pt
                  \medskipamount=7.5pt plus2pt minus2pt
                  \bigskipamount=15pt plus4pt minus4pt
                  \normalbaselineskip=15pt plus0pt minus0pt
                  \normallineskip=1pt
                  \normallineskiplimit=0pt
                  \jot=3.75pt
                  {\def\smallskip {\vskip\smallskipamount}}
                  {\def\medskip   {\vskip\medskipamount}}
                  {\def\bigskip   {\vskip\bigskipamount}}
                  {\setbox\strutbox=\hbox{\vrule
                    height10.5pt depth4.5pt width 0pt}}
                  \parskip 7.5pt
                  \normalbaselines}
\def\middlespace {\smallskipamount=5.625pt plus1.5pt minus1.5pt
                  \medskipamount=11.25pt plus3pt minus3pt
                  \bigskipamount=22.5pt plus6pt minus6pt
                  \normalbaselineskip=22.5pt plus0pt minus0pt
                  \normallineskip=1pt
                  \normallineskiplimit=0pt
                  \jot=5.625pt
                  {\def\smallskip {\vskip\smallskipamount}}
                  {\def\medskip   {\vskip\medskipamount}}
                  {\def\bigskip   {\vskip\bigskipamount}}
                  {\setbox\strutbox=\hbox{\vrule
                    height15.75pt depth6.75pt width 0pt}}
                  \parskip 11.25pt
                  \normalbaselines}
\def\doublespace {\smallskipamount=7.5pt plus2pt minus2pt
                  \medskipamount=15pt plus4pt minus4pt
                  \bigskipamount=30pt plus8pt minus8pt
                  \normalbaselineskip=30pt plus0pt minus0pt
                  \normallineskip=2pt
                  \normallineskiplimit=0pt
                  \jot=7.5pt
                  {\def\smallskip {\vskip\smallskipamount}}
                  {\def\medskip   {\vskip\medskipamount}}
                  {\def\bigskip   {\vskip\bigskipamount}}
                  {\setbox\strutbox=\hbox{\vrule
                    height21.0pt depth9.0pt width 0pt}}
                  \parskip 15.0pt
                  \normalbaselines}

\include{dspace12}
\def\be{\begin{equation}}
\def\ee{\end{equation}}
\def\bea{\begin{eqnarray}}
\def\eea{\end{eqnarray}}
\def\nn{\nonumber}

\def\ph{\phi}
\def\lt{\left}
\def\rt{\right}
\def\sect #1{\setcounter{equation}{0}}

\begin{document}
\singlespace
\begin{center}
{\LARGE{Energy and momentum of cylindrical gravitational waves. II
}}
\end{center}
\vspace{0.6in}
\begin{center}
{\large{K. S. Virbhadra \footnote[1]{Present address\ :\
Theoretical Astrophysics Group, Tata Institute of Fundamental Research,
Homi Bhabha Road, Colaba, Bombay 400005, India.
Email : shwetketu@tifrvax.tifr.res.in} \\
Physical Research Laboratory \\
Navrangpura, Ahmedabad - 380 009, India}}
\end{center}
\vspace{1.0in}
\begin{abstract}
Recently Nathan Rosen and the present author obtained  the energy and
momentum densities  of cylindrical gravitational waves in Einstein's
prescription and found them to be finite and reasonable.
In the present paper we calculate the same in prescriptions of
Tolman as well as Landau and Lifshitz and discuss the results.
\end{abstract}
\begin{flushleft}
\end{flushleft}
\vspace{1.3in}
\begin{center}
( To appear in Pramana - J. Phys. {\bf 45} 1995)
\end{center}

\middlespace
\newpage
\begin{flushleft}
{\bf 1. Introduction}
\end{flushleft}
Long ago Scheidegger $[1]$ raised doubts whether gravitational radiation
has well-defined existence. To this end, Rosen $[2]$ investigated
whether or not  cylindrical gravitational waves  have energy and momentum.
He used the energy-momentum pseudotensors of Einstein and Landau
and Lifshitz (LL) and carried out calculations in cylindrical polar
coordinates. He found that the energy and momentum density components
vanish.  The results obtained by
him fit in with the conjecture of Scheidegger that a physical system
cannot radiate gravitational energy. Two years later, Rosen
$[3]$  realized the mistake and carried out  the calculations in ``cartesian
coordinates''
and  reported that the energy and momentum densities
are nonvanishing and reasonable. Though in the letter he
remarked that he would publish the details elsewhere,
it was not done until recently he and the present
author $[4]$ re-calculated the energy and momentum density components
in ``cartesian coordinates'' in Einstein's prescription. They
found finite and reasonable results.

The physical interpretation of Einstein's energy-momentum complex
has been questioned by several physicists, notably by Weyl,
Pauli and Eddington (see in ref. $[5]$). Their objections were that the
energy-momentum complex of Einstein was  neither a tensor nor
was it  symmetric.
However, Palmer $[6]$ discussed the importance of Einstein's energy-momentum
complex in detail.
Since the first energy-momentum complex of Einstein was given, a
large number of prescriptions to obtain energy and momentum in a
general relativistic system have been proposed by many authors
(see in ref. $[7]$).
Since there is no unique way of defining energy and momentum in
a curved spacetime, and also the various energy-momentum complexes
are not tensors, some  researchers do not take them seriously.
They even guess that the different energy-momentum complexes
could give different and therefore unacceptable energy distribution
in a given spacetime.
However, we have shown that the energy-momentum pseudotensors of
Einstein, Tolman, and LL give the same result for
energy distribution  in the Kerr-Newman as well as in the Vaidya spacetimes
when calculations are carried out in ``cartesian coordinates'' $[8-10]$.
We have already mentioned that Rosen and the present author
obtained energy and momentum densities for the cylindrical gravitational
waves in ``cartesian coordinates'' in Einstein's prescription and found
reasonable result.
Therefore, it is of interest to  investigate whether or not other
energy-momentum complexes give the same result.  This is the aim of this paper.
We consider here the energy-momentum complexes of Tolman and LL.
This paper is organized as follows. In section two we write  the Einstein-
Rosen metric. In sections three and four we obtain energy and momentum
densities
of cylindrical gravitational waves in prescriptions of Tolman
and LL. In  section five we discuss the results obtained.
We use the geometrized units such that $G = 1, c = 1$ and follow
the convention that the Latin  and Greek indices take values $0$
to $3$ and $1$ to $3$, respectively. $x^0$ is the time coordinate.

\begin{flushleft}
{\bf 2. Einstein-Rosen metric}
\end{flushleft}
The Einstein-Rosen metric is a non-static vacuum solution of
Einstein's field equations and it describes the gravitational
field of cylindrical gravitational waves. It is given by the line
element $[11]$:
\be
ds^2 = e^{2\gamma - 2 \psi} \lt(dt^2-d\rho^2\rt) - e^{-2 \psi}
\rho^2 d\phi^2 - e^{2\psi} dz^2 ,
\ee

where $\gamma = \gamma(\rho,t)$, $\psi = \psi(\rho,t)$, and

\bea
\psi_{\rho \rho} &+& \frac{\psi_{\rho}}{\rho} - \psi_{tt} = 0 , \\
\gamma_{\rho} &=& \rho \left({\psi}_{\rho}^{2} + \psi_t^2\right), \\
\gamma_t &=& 2\ \rho\ \psi_{\rho}\ \psi_t \ .
\eea

The subscripts on $\psi$ and $\gamma$ denote partial derivatives
with respect to the subscripts. It is known that the
energy-momentum complexes of Tolman and LL, like that of
Einstein, give correct result if calculations are carried out
in ``cartesian coordinates''[12-17]. Therefore, one transforms the line
element,
given by $(1)$, according to
\bea
x = \rho\ cos\ph , \nn\\
y = \rho\ sin\ph ,
\eea
and gets the   line element in $t,x,y,z$ coordinates $[4]$,
\be
ds^2 = e^{2(\gamma-\psi)}\ \left[dt^2 - \frac{(x dx + y dy)^2}{\rho^2}
      \right] - \frac{e^{-2\psi}}{\rho^2} {\left(x dy - y dx\right)}^2
      - e^{2\psi} dz^2 . \\
\ee

\begin{flushleft}
{\bf 3. Energy and momentum of cylindrical gravitational waves
in prescription of Tolman}
\end{flushleft}
The energy-momentum complex of Tolman is $[13-14]$
\be
{\cal{T}}^i_k = \frac{1}{8 \pi} \ U^{ij}_{k,\ j} ,
\ee
where
\bea
U^{ij}_k &=& \sqrt{-g} \lt[
-g^{li}\ \lt(-\Gamma^j_{kl}\ +\ \frac{1}{2}\ g^j_k\ \Gamma^{a}_{al}\
                             +\ \frac{1}{2}\ g^j_l\  \Gamma^{a}_{ak}\rt)
                             \right.\nn\\
&+&\left.
\ \frac{1}{2}\ g^i_k\  g^{lm}\ \lt(-\Gamma^j_{lm}\ +\ \frac{1}{2}\ g^j_l\
\Gamma^a_{am}\ +\ \frac{1}{2}\ g^j_m\ \Gamma^a_{al}\rt)
                      \rt] \ .
\eea
${\cal{T}}^0_0$ is the energy density, ${\cal{T}}^0_{\alpha}$ are the momentum
density components, and ${\cal{T}}^{\alpha}_0$ are the components of
energy current density.
The required nonvanishing components of $U^{ij}_k$  are
\bea
U^{01}_0 &=& \frac{x \lt( e^{2\gamma} - 1\rt)}{2 \rho^2},\nn\\
U^{02}_0 &=& \frac{y \lt( e^{2\gamma} - 1\rt)}{2 \rho^2},\nn\\
U^{01}_1 &=& \frac{- 2 \ \psi_t\ \psi_{\rho}\ y^2}{\rho},\nn\\
U^{02}_2 &=& \frac{- 2 \ \psi_t\ \psi_{\rho}\ x^2}{\rho},\nn\\
U^{02}_1 &=& U^{01}_2 = \frac{2 \ \psi_t \ \psi_{\rho} x y} {\rho},\nn\\
U^{03}_3 &=& 2\ \psi_t\ \lt( 1 - \rho\ \psi_{\rho}\rt),\nn\\
U^{11}_0 &=& \frac{2 e^{2\gamma}\ \psi_t \ \psi_{\rho}\ y^2}{\rho},\nn\\
U^{22}_0 &=& \frac{2 e^{2\gamma}\ \psi_t \ \psi_{\rho}\ x^2}{\rho},\nn\\
U^{12}_0 &=& U^{21}_0 = \frac{ -2 e^{2\gamma}\ \psi_t\ \psi_{\rho} x y}
                           {\rho},\nn\\
U^{33}_0 &=& 2 e^{2 \gamma - 4 \psi}\ \psi_t \ \lt(\rho\
           \psi_{\rho} - 1\rt) \ .
\eea
Using $(9)$ in $(7)$ we obtain
\bea
{\cal{T}}^0_0 &=& \frac{e^{2 \gamma}\lt({\psi_{\rho}}^2 +
{\psi_t}^2\rt)}{8 \ \pi},\nn\\
{\cal{T}}^0_1 &=& \frac{x\ \psi_{\rho}\ \psi_t}{4\pi\rho},\nn\\
{\cal{T}}^0_2 &=& \frac{y\ \psi_{\rho}\ \psi_t}{4\pi\rho},\nn\\
{\cal{T}}^1_0 &=& - e^{2\gamma}\ {\cal{T}}^0_1,\nn\\
{\cal{T}}^2_0 &=& - e^{2\gamma}\ {\cal{T}}^0_2,\nn\\
{\cal{T}}^3_0 &=& {\cal{T}}^0_3 = 0 \ .
\eea

\begin{flushleft}
{\bf 4. Energy and momentum of cylindrical gravitational waves
in prescription of LL}
\end{flushleft}
The energy-momentum complex of LL is [15]
\be
L^{mn} = \frac{1}{16 \pi} {S^{mjnk}}_{,jk} ,
\ee
where
\be
S^{mjnk} =  -g \left( g^{mn} g^{jk} - g^{mk} g^{jn}\right) .
\ee
$L^{mn}$ is symmetric in its indices. $L^{00}$ is the energy
density and $L^{0\alpha}$ are the momentum (energy current) density components.
$S^{mjnk}$ has symmetries
of the Reimann curvature tensor.The required nonvanishing components of
 $S^{mjnk}$ are
\bea
S^{0101} &=& \frac{-\left(e^{2\gamma}\ y^2 +\ x^2\right)}{\rho^2},\nn\\
S^{0102} &=& \frac{x\ y\ \left( e^{2\gamma} -1 \right)}{\rho^2},\nn\\
S^{0202} &=& \frac{-\left(e^{2\gamma} x^2 + y^2\right)}{\rho^2},\nn\\
S^{0303} &=& - e^{2\left(\gamma - 2 \psi\right)}\ .
\eea
Making use of $(13)$ in $(11)$ we obtain energy and momentum (energy
current) density components
for the cylindrical gravitational waves in LL prescription.
\bea
L^{00} &=& \frac{e^{2\gamma}\left({\psi_{\rho}}^2 + {\psi_t}^2 \right)}
         {8\ \pi},\nn\\
L^{01} &=& \frac{- e^{2 \gamma}\  \psi_{\rho}\ \psi_t\ x}
              {4\ \pi\ \rho},\nn\\
L^{02} &=& \frac{- e^{2 \gamma}\  \psi_{\rho}\ \psi_t\ y}
              {4\ \pi\ \rho},\nn\\
L^{03} &=&  0 \ .
\eea

\begin{flushleft}
{\bf 5. Discussion}
\end{flushleft}
One can see from  $(10)$ and $(14)$ that the energy-momentum
complexes of Tolman and LL give the same energy and energy current
densities as given by Einstein's prescription $[4]$.
The momentum density components in the Tolman prescription are the
same as in the Einstein prescription.
The  momentum density components in Tolman and LL
prescriptions differ by a sign, as expected.
The energy density of the cylindrical gravitational
waves is finite and positive definite, and the momentum density
components reflect the symmetry of the spacetime. Thus our
investigations do not support the conjecture of Scheidegger
$[1]$.

Due to non-tensorial nature of  energy-momentum complexes,
some physicists do not take them seriously. They even
conjecture that different energy-momentum complexes could give
different energy distributions in a given spacetime. To this end
we have shown that the well-known energy-momentum complexes of
Einstein, Tolman, and LL give the same and reasonable  result for the
Kerr-Newman, Vaidya, and Einstein-Rosen spacetimes (see also
references $4,7-10$).

\begin{flushleft}
{\bf Note added in proof}\\
\end{flushleft}
Using the Papapetrou pseudotensor we have obtained the energy
and energy current (momentum) densities of the cylindrical gravitational
waves (described by the Einstein-Rosen metric) and have found
them to be the same as given by the LL energy-momentum complex.

\begin{flushleft}
{\bf Acknowledgements}\\
\end{flushleft}
Thanks are due to  P. C. Vaidya, Nathan Rosen, and Alberto
Chamorro for careful reading of the manuscript, and to the Basque Government
and the Tata Institute of Fundamental Research for partial  support for
completion
of this work.
Thanks are also due to the referee for suggestions.

\newpage
\begin{flushleft}
{\bf References}
\end{flushleft}
1. A E Scheidegger, {\it  Rev. Mod. Phys. } {\bf 25} \  451\ (1953)\\
2. N  Rosen, {\it  Helv. Phys. Acta. Suppl.} {\bf 4}  \ 171 \ (1956)\\
3. N Rosen, {\it   Phys. Rev. } {\bf 110}\  291\ (1958)\\
4. N Rosen and K S Virbhadra, {\it  Gen. Relativ.. Gravit.} {\bf 26} \ 429 \
(1993)\\
5. S Chandrasekhar and V  Ferrari, {\it   Proc. R. Soc. Lond. A } {\bf 435} \
645 \
  (1991)\\
6. T Palmer, {\it Gen. Relativ.. Gravit. } {\bf 12} \ 149 \ (1980) \\
7. K S Virbhadra, {\it   Math. Today } {\bf 9} \ 39 \ (1991)\\
8. K S Virbhadra, {\it   Phys. Rev. } {\bf D41} \ 1086 \ (1990)\\
9. K S Virbhadra, {\it   Phys. Rev. } {\bf D42} \ 2919\ (1990)\\
10. K S Virbhadra, {\it   Pramana-J. Phys. } {\bf 38} \ 31 \ (1992)\\
11. M Carmeli, {\it   Classical Fields : General Relativity
and Gauge Theory} (John Wiley and Sons, 1982)\ p.198\\
12. C M\o ller, {\it  Ann. Phys. (NY)} {\bf 4} \ 347\ (1958)\\
13. P C  Vaidya {\it J. Univ. Bombay } {\bf 21} \  \ 3 \ 1\ (1952)\\
14. R C Tolman, {\it  Relativity  Thermodynamics and
Cosmology} (Oxford Univ. Press  London,  1934)  p.227\\
15. L D  Landau and E M Lifshitz, {\it   The Classical
Theory of Fields} (Pergamon Press, 1987) p.280\\
16. R W Lindquist, R A  Schwartz, and C W Misner, {\it Phys. Rev. } {\bf 137}
  \ B1364 (1965) \\
17. K S Virbhadra  and J C Parikh, {\it  Phys. Lett. } {\bf B317} \ 312 \
(1993)\\
\end{document}